\documentclass[9pt,twocolumn,twoside]{osajnl}

\journal{ol} 
\setboolean{shortarticle}{true} 

\title{Control of Second Harmonic Generation in Doubly Resonant Aluminum Nitride Microrings to Address Rubidium Two-Photon Clock Transition}

\author[1]{Joshua B. Surya}
\author[1]{Xiang Guo}
\author[1,2]{Chang-Ling Zou}
\author[1,*]{Hong X. Tang}

\affil[1]{Department of Electrical Engineering, Yale University, New Haven, Connecticut 06511, USA}
\affil[2]{Department of Optics and Optics Engineering, University of Science and Technology of China, Hefei, Anhui 230026, China}

\affil[*]{Corresponding author: hong.tang@yale.edu}

\ociscodes{(190.0190) Nonlinear optics; (190.2620) Harmonic generation and mixing; (130.0130) Integrated optics.}

\begin{abstract}
Nonlinear optical effects have been studied extensively in microresonators as more photonics applications transition to integrated on-chip platforms.  Due to low optical losses and small mode volumes, microresonators are demonstrably the state-of-the-art platform for second harmonic generation (SHG).  However, the working bandwidth of such microresonator-based devices are relatively small, presenting a challenge for applications where a specifically targeted wavelength needs to be addressed. In this work, we analyzed the phase-matching window and resonance wavelength with respect to varying microring width, radius and temperature. A chip with precise design parameters was fabricated with phase-matching realized at the exact wavelength of two-photon transition of 85-Rubidium.  This procedure can be generalized to any target pump wavelength in the telecom-band with picometer precision.  
\end{abstract}

\begin{document}

\maketitle


Second harmonic generation (SHG) has been studied extensively throughout a variety of systems, with applications ranging from molecular imaging \cite{Han2005,Moreaux2000,Cox2003}, quantum optics \cite{Kwiat1995,Bouwmeester1997,Guo2016a} to ultrafast optics \cite{Abraham2014,Petersen2010,Weiner2009} and many others \cite{Wehling2015,Yu2002,Udem2002,Cazzanelli2011,Glazov2011,Kruk2015}.  In addition to having low optical losses, the compact and scalable nature of on-chip devices pose a significant advantage over other platforms.  In hopes of achieving the maximum conversion efficiency possible, integrated microring resonators have been used to enhance the SHG output \cite{Klein2006,Niesler2009,Pernice2012,Sakai2006,Yu2002,Levy2011,Yang2007}.  Very high efficiencies of up to 2500$\,\%$/W have been observed previously \cite{Guo2016}, but due to the high quality (Q) factor and phase-matching conditions of these systems, the usable bandwidth is often limited to <0.1nm. Precise tuning of optimized SHG at wavelengths of interests remains to be a major challenge of integrated on-chip platforms due to the demanding requirements on nanofabrication process variations.

Frequency doubling of the micro-frequency comb \cite{Udem2002,Diddams2010} has been of particular interest since the demonstration of broadband comb generation on microresonator platforms \cite{Diddams2010,Brasch:17,DelHaye2007,Hong:03,Jung:13,Luke:15,Okawachi2011}.  These combs are useful when the carrier envelope offset frequency is stabilized, where efficient SHG of a specific comb line is vital to the process\cite{Brasch:17,DelHaye2016}.  Although comb line frequency doubling has been demonstrated using quasi-phase-matching in periodically poled lithium niobate \cite{Brasch:17,Sakai2006}, this process is performed off-chip.  By integrating the comb-generation and doubling processes on the same chip, much higher comb locking efficiencies and minimal device footprint are expected. In addition, by locking a SH signal of a comb line to a stable atomic transition in the optical frequencies using microresonators, it is possible to create compact clocks that are orders of magnitude more stable than microwave based Cesium time standards \cite{Diddams2001,Diddams2004}. All of these applications demand the precise tuning of the SHG wavelength in our microresonator devices with picometer precision.

\begin{figure}[t]
\centering
\includegraphics[width=\linewidth]{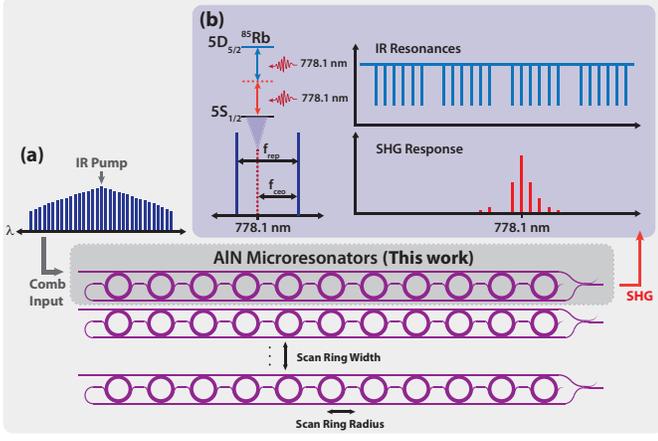}
\caption{(a) A chip consisting of arrays of AlN microrings multiplexed in each individual bus waveguides. For optical clock application, an IR frequency comb can serve as an input to these devices with optimized SHG for f-2f self-referencing interferometry. (b) One device with systematically designed microring parameters was used for optimal phase-matching of the fundamental IR mode at $1556.24\,\mathrm{nm}$ with the high order SH $\mathrm{TM}_2$ mode at $778.12\,\mathrm{nm}$, corresponding to the two-photon transition of $^{85}\mathrm{Rb}$. This is an important step towards simultaneous comb-locking and stable optical frequency reference using atomic transitions.}
\label{fig1:motivation}
\end{figure}

In this work, we demonstrate that by systematically varying the cross sectional geometry and ring radius of the microring resonators on a polycrystalline aluminum nitride (AlN) nonlinear photonics platform as well as controlling the system temperature, highly efficient SHG can be achieved at a specific wavelength of interest.  Here we realize the phase-matching wavelength to address the two photon transition of $^{85}\mathrm{Rb}$ at $\lambda_{\mathrm{Rb}} = 778.12\,\mathrm{nm}$ \cite{Edwards2005}.  Our devices utilize an array of microrings with optimized Q factor, coupling conditions and cross-sectional geometry for the systematic analysis of the phase-matching conditions and resonance locations. These preliminary devices guided the design of our final chip which experimentally realized high efficiency SHG output at $778.1\,\mathrm{nm}$ with both pump and SH modes in resonance. The on-chip efficiency achieved was 1800$\%$/W.

In order to maximize yield, several polycrystalline AlN chips consisting of 8 or more microring resonators were fabricated for preliminary analysis of the optimal device parameters. The advantages of the AlN platform are numerous. First, it can be deposited by sputtering at low temperatures, allowing a smooth surface and uniform thickness across the entire wafer, leading to consistent results between multiple fabrication runs. AlN has also been demonstrated to have large Pockels ($x^{(2)}$) effect \cite{Jung2014,Jung2014a,Jung2016}, with SHG efficiency as high as 2500$\,\%$/W having been observed \cite{Guo2016}.  Additionally, microrings with small radii can easily be realized on AlN using conventional fabrication techniques, making this platform feasible for large-scale production. Finally, AlN has excellent optical properties with a large bandgap, exhibiting low loss at visible wavelengths.

There are two design challenges that must be addressed for highly efficient SHG at a target wavelength. (1) The efficient coupling to both fundamental and SH wavelengths, and (2) controlling of the resonant modes that simultaneously match the desired fundamental and SH wavelengths in a single cavity. For example, to realize an optical clock network based on fiber-optics, a telecom laser with a wavelength of $1556.24\,\rm{nm}$ can be locked to the two-photon transition in $^{85}\mathrm{Rb}$ at $\lambda_{\mathrm{Rb}}$ by SHG, where a microring with two resonances tuned to the exact IR and visible wavelengths is highly useful. This concept is illustrated in Fig.$\,\ref{fig1:motivation}$(a) and (b). The first challenge has been resolved by the authors \cite{Guo2016} by introducing two loading waveguides to couple the two wavelengths independently. 

\begin{figure}[ht]
\centering
\includegraphics[width=\linewidth]{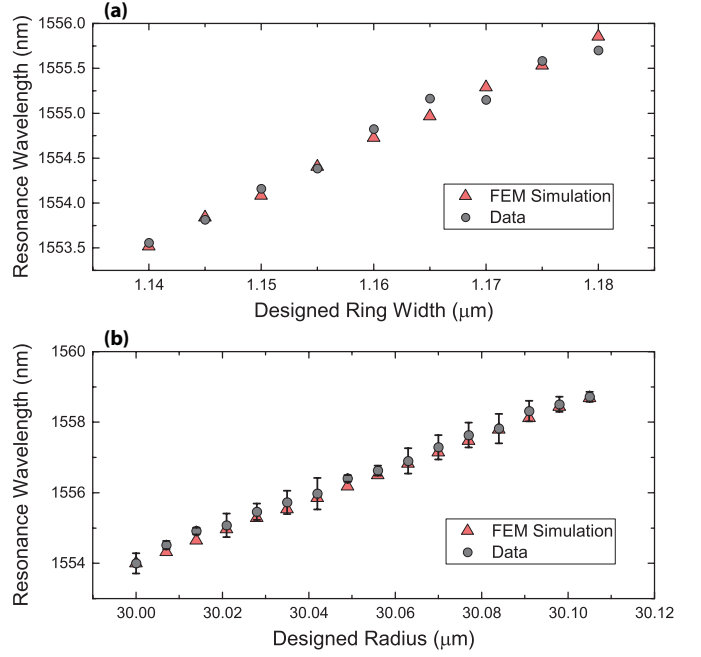}
\caption{(a) Analysis of the phase-matching condition with varying width. In order for our simulation to be as accurate as possible, the cross section of our ring includes details of the cross section of the fabricated ring, including a sidewall angle of $8$ degrees.  The refractive index curve of our material was also carefully selected.  The experimental data matches with the simulated data reasonably well.  (b) Analysis of the resonance wavelength at varying radii.}
\label{fig2:widthradiusvary}
\end{figure}
\begin{figure}[t]
\centering
\includegraphics[width=\linewidth]{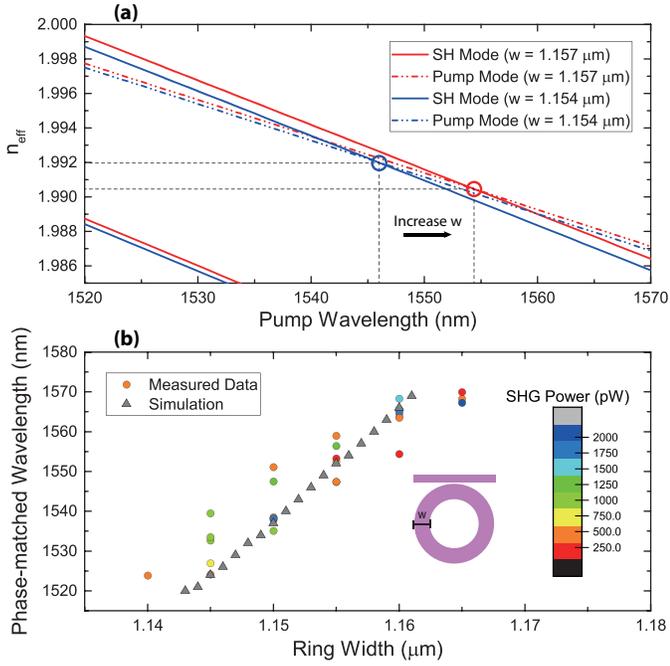}
\caption{(a) The dependence of the phase-matching wavelength with increasing ring width.  Phase-matching occurs at larger pump wavelengths with wider ring widths. (b) Comparison of simulated and measured phase-matched wavelength with designed ring width.  Measured data is color coded by SHG output power.}
\label{fig3:widthvary}
\end{figure}

The second challenge originates from the fact that the wavelengths of modes in a single mciroring cannot be controlled independently. To address this challenge, we first vary the width to coarsely adjust the phase-matching window to $\lambda_{\mathrm{Rb}}$, after which multiple geometry parameters as well as the temperature were adjusted to realize independent control of the resonance frequencies of the two modes. From a theoretical standpoint, for a given fundamental IR frequency orbital angular momentum (OAM) number $m$, conservation of momentum requires the SH mode to have twice the OAM $2m$. In general, for a bent waveguide with width $w$ and radius $R$, the effective index $n_{\mathrm{eff}}^{\left(1\mathrm{f}\right)}\left(n_{\mathrm{eff}}^{\left(2\mathrm{f}\right)}\right)$ of a mode at wavelength $2\lambda_{\mathrm{Rb}}\left(\lambda_{\mathrm{Rb}}\right)$ and OAM $m\left(2m\right)$follows the equation 
\begin{align}
\left(n_{\mathrm{eff}}^{\left(q\mathrm{f}\right)}+\frac{\partial n_{\mathrm{eff}}^{\left(q\mathrm{f}\right)}}{\partial\lambda}\delta\lambda^{\left(q\mathrm{f}\right)}\right)2\pi R & =m\left(q\lambda_{\mathrm{Rb}}+\delta\lambda^{\left(q\mathrm{f}\right)}\right),
\end{align}
with $q\in\{1,2\}$ denoting fundamental and SH mode. If we take the ring width $w$, ring radius $R$ and operating temperature of the microring $T$ as independent parameters to control the mode frequency, we arrive at the variation relation 
\begin{align*}
d\lambda^{\left(q\mathrm{f}\right)} & \approx A^{\left(q\mathrm{f}\right)}dw+B^{\left(q\mathrm{f}\right)}dT+C^{\left(q\mathrm{f}\right)}dR+D^{\left(q\mathrm{f}\right)}dm,
\end{align*}
with $A^{\left(q\mathrm{f}\right)}=\frac{1}{n_{\mathrm{g}}^{\left(q\mathrm{f}\right)}}\frac{\partial n_{\mathrm{eff}}^{\left(q\mathrm{f}\right)}}{\partial w}$,
$B^{\left(q\mathrm{f}\right)}=\frac{1}{n_{\mathrm{g}}^{\left(q\mathrm{f}\right)}}\frac{\partial n_{\mathrm{eff}}^{\left(q\mathrm{f}\right)}}{\partial T}$,
$C^{\left(q\mathrm{f}\right)}=\frac{\pi qm\left(n_{\mathrm{eff}}^{\left(q\mathrm{f}\right)}-\frac{\partial n_{\mathrm{eff}}^{\left(q\mathrm{f}\right)}}{\partial\lambda}\right)}{\left[n_{\mathrm{g}}^{\left(q\mathrm{f}\right)}\right]^{2}2\pi R^{2}}$
and $D^{\left(q\mathrm{f}\right)}=\frac{qn_{\mathrm{eff}}^{\left(q\mathrm{f}\right)}}{\left[n_{\mathrm{g}}^{\left(q\mathrm{f}\right)}\right]^{2}2\pi R}dm$.
From our empirical observations, $n_{\mathrm{g}}^{\left(q\mathrm{f}\right)}=\frac{qm}{2\pi R}-\frac{\partial n_{\mathrm{eff}}^{\left(q\mathrm{f}\right)}}{\partial\lambda}\gg1$, thus the coefficients $A,B,C,D$ can be treated as constants. From the experimental results illustrated in Fig.$\,\ref{fig2:widthradiusvary}$(a) and (b), we find the resonance wavelength shift coefficient to be $A^{\left(1\mathrm{f}\right)}\approx0.055\pm0.0028$ and $C^{\left(1\mathrm{f}\right)}\approx0.043\pm0.00053$. The resonance of the microring resonator with different radius and width are simulated by COMSOL, considering our microring as a waveguide cross-section with an axis-symmetric condition.  By setting the mode number to be 226, we simulate the eigen-frequency (as well as resonance wavelength since $\lambda_{res} = c/f_{res}$) at different microring radius and waveguide width.  The simulation result is shown in Fig.$\,\ref{fig2:widthradiusvary}$, which matches well with the experimental data. The thermo-shift coefficient of the infrared (IR) resonance was measured to be $B^{\left(1\mathrm{f}\right)}=0.019\,\pm\,0.00031\,\mathrm{nm}/\mathrm{K}$. $D^{\left(1\mathrm{f}\right)}$ can be estimated to be about $5.84$, corresponding to the FSR of the modes. The phase-matching condition also requires $d\lambda^{\left(1\mathrm{f}\right)}=2d\lambda^{\left(2\mathrm{f}\right)}$. Therefore, for fixed $R$ and $T$, if $m$ is changed by $1$, the corresponding phase-matching wavelength shifts by $D^{\left(\mathrm{1f}\right)}-A^{\left(\mathrm{f}\right)}\frac{D^{\left(1\mathrm{f}\right)}-2D^{\left(2\mathrm{f}\right)}}{A^{\left(1\mathrm{f}\right)}-2A^{\left(2\mathrm{f}\right)}}$, with $dw=-\frac{D^{\left(1\mathrm{f}\right)}-2D^{\left(2\mathrm{f}\right)}}{A^{\left(1\mathrm{f}\right)}-2A^{\left(2\mathrm{f}\right)}}$. The rate of shift in $dw$ is $\sim2.8\pm0.026$ {[}as illustrated in Fig.$\,$\ref{fig3:widthvary}(a) and (b){]}, which is relatively large due to the Vernier effect. We select the mode close to $2\lambda_{\mathrm{Rb}}$ (i.e. fixed $m$) with a maximum deviation of $\frac{1}{2}D^{\left(1\mathrm{f}\right)}$, while varying the parameters $w$, $R$ and $T$ in order to align the IR resonance to $2\lambda_{\mathrm{Rb}}$. From the coefficients, we found that to tune the resonance by $\frac{1}{2}D^{\left(1\mathrm{f}\right)}$, the corresponding change in width, temperature or radius is, $9D^{(1\mathrm{f})}\,\mathrm{nm}$, $26D^{(\mathrm{1f})}\,\mathrm{K}$, or $11.5D^{(1\mathrm{f})}\,\mathrm{nm}$ respectively. Meanwhile, the phase-matching condition still requires $\left(A^{\left(\mathrm{f}\right)}-2A^{\left(2\mathrm{f}\right)}\right)dw+\left(B^{\left(\mathrm{f}\right)}-2B^{\left(2\mathrm{f}\right)}\right)dT+\left(C^{\left(\mathrm{f}\right)}-2C^{\left(2\mathrm{f}\right)}\right)dR=0$ to be satisfied. 

Guided by the preceding analysis, we control $w$, $T$, and $R$ to optimize the SHG output such that $\lambda^{\left(2\mathrm{f}\right)}=\lambda_{\mathrm{Rb}}$($d\lambda^{\left(2\mathrm{f}\right)}=0$). Since $w$ and $R$ are fixed by fabrication, and $T$ is re-configurable in real-time with relatively small thermo-shift coefficients $(B^{\left(1\mathrm{f}\right)},B^{\left(2\mathrm{f}\right)})$, thermal tuning was used as a fine-tuning approach whereas the geometry parameters $w$ and $R$ were varied for coarse tuning.

For the preliminary chip designs, the width of the microrings was varied from $1.13-1.18\,\mu\mathrm{m}$ at intervals of $10\,$nm in order to extract the approximate range of ring widths that phase-matched to the working range of our tunable telecom-band laser at $1520-1570\,\mathrm{nm}$. By analyzing the data from three separate chips, the phase-matching width was found to be $1.13-1.17\,\mu\mathrm{m}$. Guided by this result, the ring width for the final chip fabricated was scanned at narrower ranges with a fine gap of $5\,\mathrm{nm}$. With a SHG output wavelength bandwidth of $\sim7\,\mathrm{nm}$ for each array of microrings, phase-matching can be realized for any single wavelength within the bandwidth of the laser in every fabricated chip. 

To maximize yield and determine appropriate scanning ranges for our first few chips, we designed each device to consist of a cascade of 16 rings as shown in Fig.\,\ref{fig1:motivation}, where the radius is scanned with a step of 7$\,\mathrm{nm}$. As shown in Fig.$\,$\ref{fig2:widthradiusvary}(b), each radius step corresponded to a $36\,\pm\,3.7\,\mathrm{pm}$ shift of mode wavelength, effectively covering a span of $5.7\,\mathrm{nm}$, with a radius change of $100\,nm$. In the final chip design, the sweeping range of ring width further narrowed to $1.15-1.17\,\mu\mathrm{m}$. The final working device had a designed ring width of $1.155\,\mu\mathrm{m}$ and a radius of $30.035\,\mu\mathrm{m}$.

As shown from numerous previous studies\cite{Surya:18,Guo2016}, doubly-resonant SHG efficiency $\eta$ in the non depletion regime is 
\begin{equation}
\eta=\frac{\mathrm{P}_{\mathrm{SH}}}{\mathrm{P}_{\mathrm{p}}^{2}}=g^{2}\frac{4\kappa_{\mathrm{SH},1}}{\delta_{\mathrm{SH}}^{2}+(\kappa_{\mathrm{SH}})^{2}}\Bigg(\frac{2\kappa_{\mathrm{p},1}}{\delta_{\mathrm{p}}^{2}+(\kappa_{\mathrm{p}})^{2}}\Bigg)^{2},\label{eq2:shgefficiency}
\end{equation}
where $g$ is the nonlinear coupling rate, $\kappa_{\mathrm{SH}(\mathrm{p}),0}$ and $\kappa_{\mathrm{SH}(\mathrm{p}),1}$ are the SH (pump) mode intrinsic loss and external coupling rates, respectively.

The following requirements described by Eq.$\,\ref{eq2:shgefficiency}$ must be fulfilled in order to optimize phase-matching and SHG output at $\lambda_{\mathrm{Rb}}$. (1) IR coupling: the comb line power is normally on the order of $-20\,\mathrm{dbm}$, thus the AlN ring will typically be operating at the non-depletion regime. Accordingly, the system should be critically coupled for both visible and IR wavelengths. A straight bus waveguide of $0.57\,\mu\mathrm{m}$ width was used for the coupling of IR light, and is tapered to $3\,\mu\mathrm{m}$ to facilitate the overall transmission of the device. (2) Visible light coupling: since the extraction of SH light through the IR bus is inefficient, a wrap-around waveguide tapered from $0.17-0.12\,\mu\mathrm{m}$ was used to optimize the extraction of visible light. This waveguide was bent around to guide the SH light towards the output end of the chip (shown in Fig.$\,$\ref{fig1:motivation}(a)). Furthermore, an integrated wavelength division multiplexer was used to combine IR and Visible light onto the same waveguide. (3) By fine tuning using temperature control, the detuning of the SH light and IR pump input $\delta_{SH(\mathrm{p})}$ is set to zero.

\begin{figure}[t]
\centering
\includegraphics[width=\linewidth]{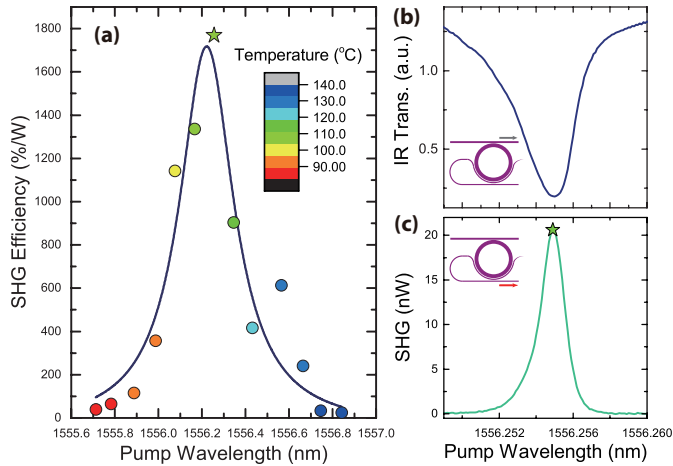}
\caption{(a) Temperature dependent SH response on a wavelength scale and a converted temperature scale using the thermo-shift coefficient of polycrystalline AlN at IR wavelengths.  A maximum internal efficiency of $1800\,\%/W$was observed at a wavelength of $1556.2\,\mathrm{nm}$ ($110\,^\circ\mathrm{C}$) with a tuning bandwidth (FWHM) of $0.303\,\pm\,0.068\,\mathrm{nm}$ ($17.5\, ^\circ \mathrm{C}$).  Each data point is color coded by its corresponding temperature.  (b) A zoomed-in spectrum of the transmission at $110\,^\circ\mathrm{C}$. (c) SHG response at $110\,^\circ\mathrm{C}$, with the peak response at the visible wavelength corresponding to $\lambda_{\mathrm{Rb}}$.}
\label{fig4:results}
\end{figure}

Due to the thermal heating properties inherent of polycrystalline AlN, at varying temperatures the IR and visible resonances will have different shifts in wavelength \cite{Guo2016,Surya:18}. The effect of these thermal effects on SHG power is shown in Fig.$\,$\ref{fig4:results}(a) and is modeled by the following equation, 
\begin{equation}
\eta\approx\frac{\hbar\omega_{\mathrm{SH}}}{(\hbar\omega_{\mathrm{p}})^{2}}\frac{8g^{2}\kappa_{\mathrm{SH},1}\kappa_{\mathrm{p},1}^{2}}{\kappa_{\mathrm{p}}^{4}\big[\frac{2\pi c}{\lambda_{\mathrm{p}}\lambda_{\mathrm{SH}}}(\lambda_{\Delta}-d_{\Delta}T)^{2}+\kappa_{\mathrm{SH}}^{2}\big]}.\label{eq:shgtemp}
\end{equation}
Where the wavelength mismatch in the visible and IR resonances is compensated by the differential thermo-shift coefficients $d_{\Delta}$ between the SH and pump modes at temperature $T$. At the temperature where $\lambda_{\Delta}-d_{\Delta}T=0$, maximum SHG power can be observed. For our devices, this temperature was at $110\,^{\circ}\mathrm{C}$ with a full-width at half maximum (FWHM) of $\Delta T=17.5\,^{\circ}\mathrm{C}$. The FWHM in the wavelength scale is observed to be $0.303\,\pm\,0.068\mathrm{nm}$. At $110\,^{\circ}\mathrm{C}$, the measured IR transmission spectrum and visible response can be seen in Fig.\,\ref{fig4:results}(b). For the IR resonance, a loaded Q $\approx3\times10^{5}$ was measured with an extinction ratio of $0.8$. The maximum SH response observed was $20\,\mathrm{nW}$ at a center wavelength of $1556.255\,\mathrm{nm}$ and a FWHM of $1.5\,\mathrm{pm}$, corresponding to a maximum on-chip efficiency of 1800$\,\%/\mathrm{W}$. The input power of the pump laser was set to $180\,\mu\mathrm{W}$.

In conclusion, inspired by the numerous applications that would benefit from highly efficient SHG at a specific wavelength, we performed systematic simulations and theoretical analysis on the effect of the geometry of the microring on phase-matching. Together with data analysis from multiple chip designs, we realized a robust method of designing a phase-matched second-harmonic generator for any target fundamental wavelength in the telecom-band. As a proof of principle for applications in ultra-stable clock designs, we were able to optimize SHG output at the wavelength of $778.12\,\mathrm{nm}$, corresponding to the two-photon transition frequency of $^{85}\mathrm{Rb}$. The systematic approach is useful for designs where doubly- or triply-resonant conditions are required. This work can be generalized and extended to other on-chip material platforms for frequency conversion at targeted wavelengths.

\section*{Funding Information}
This work was conducted as part of the Draper-NIST collaboration on the DARPA ACES program under contract HR0011-16-C-0118. The authors acknowledge discussions with all the team members supported under this program. H.X.T. acknowledges support from David and Lucile Packard Foundation.

\section*{Acknowledgments}
Facilities used for device fabrication were supported by Yale SEAS cleanroom and Yale Institute for Nanoscience and Quantum Engineering. The authors thank Michael Power and Dr. Michael Rooks for assistance in device fabrication. 
\bigskip
\noindent

\bibliography{shgpaper}

\bibliographyfullrefs{shgpaper}
 

\end{document}